\documentclass{ppig}

\usepackage{titlesec}

\titlespacing*{\subsection}{0pt}{0.5em}{0em}

\usepackage{epsfig}
\usepackage{booktabs}
\usepackage[utf8]{inputenc}

\usepackage{natbib}
\usepackage[toc,page]{appendix}
\usepackage{longtable}
\usepackage{xcolor}
\usepackage[colorlinks,citecolor=teal,urlcolor=gray]{hyperref}
\usepackage{lscape} %

\usepackage[absolute,overlay]{textpos}
\usepackage{alltt}

\title{Vibe coding: programming through conversation with artificial intelligence}

\author{Advait Sarkar \\
    University of Cambridge and \\University College London \\
    United Kingdom \\
    advait.sarkar@cl.cam.ac.uk \\
    \And
    Ian Drosos \\
    Microsoft Research \\
    United Kingdom \\
    t-iandrosos@microsoft.com \\}
  
\date{}

\begin{document}

\begin{textblock*}{\dimexpr\paperwidth-2cm\relax}(1cm,0.5cm) %
  \fbox{\parbox{\dimexpr\paperwidth-2cm\relax}{\footnotesize
  \color{darkgray}
    \textbf{Citation}: Sarkar, Advait, and Drosos, Ian. 2025. ``Vibe coding: programming through conversation with artificial intelligence.'' In Proceedings of the 36th Annual Conference of the Psychology of Programming Interest Group (PPIG 2025).
\textbf{BibTeX}: \detokenize{@inproceedings{sarkar2025vibecoding, 
    author={Sarkar, Advait and Drosos, Ian}, 
    booktitle={{Proceedings of the 36th Annual Conference of the Psychology of Programming Interest Group (PPIG 2025)}},
    title={{Vibe coding: programming through conversation with artificial intelligence}}, 
    month=sep,
    year={2025}}
    }
}}
\end{textblock*}

\maketitle

\pagestyle{plain}

\begin{table}[h!]
\centering
\footnotesize
\renewcommand{\arraystretch}{1.2}
\begin{tabular}{|p{0.15\linewidth} p{0.8\linewidth}|}
\hline
\textbf{Category} & \textbf{Associated questions} \\
\hline
Goals & What kinds of applications are vibe coded? What degree of success is expected? To what extent is exploration a motivation? (Section~\ref{sec:results-goals})\\
Intentions & How are implementation ideas initially formed and refined? Is intent pre-defined or shaped iteratively through interaction with AI? (Section~\ref{sec:results-intentions})\\
Workflow & What are the stages of the vibe coding workflow? How is programmer effort distributed? How are various tools and resources combined? (Section~\ref{sec:results-workflow})\\
Prompting & How do programmers express their intent to the system? What setup and prompting strategies are used? What is the granularity of instructions? Are prompts single- or multi-objective? What input modes are employed? (Section~\ref{sec:results-prompting})\\
Debugging & What methods are used to identify and resolve problems in AI-generated code? (Section~\ref{sec:debugging}) \\
Challenges & What technical, conceptual, and workflow-related obstacles arise in vibe coding beyond debugging? What strategies are used to overcome them? (Section~\ref{sec:results-challenges})\\
Expertise & What forms of knowledge are deployed in vibe coding? When do practitioners transition from instructing AI to direct manual work? (Section~\ref{sec:results-expertise})\\
Trust & How do users build confidence in AI outputs? What are the risks of overreliance? (Section~\ref{sec:results-trust})\\
Definition and Performance & How is vibe coding understood relative to other AI-assisted programming practices? How do performative contexts such as video streaming shape the practice and its perception? (Section~\ref{sec:results-definition-performance})\\
\hline
\end{tabular}
\caption{Major categories of findings from our study of the practice of vibe coding. Findings were derived from in-depth qualitative framework analysis of approximately 8.5 hours of video of extended think-aloud vibe coding sessions.}
\label{tab:framework-overview}
\end{table}

\begin{abstract}
We examine ``vibe coding'': an emerging programming paradigm where developers primarily write code by interacting with code-generating large language models rather than writing code directly. We present the first empirical study of vibe coding. We analysed over 8 hours of curated video capturing extended vibe coding sessions with rich think-aloud reflections. Using framework analysis, we investigated programmers' goals, workflows, prompting techniques, debugging approaches, and challenges encountered.

We find that vibe coding follows iterative goal satisfaction cycles where developers alternate between prompting AI, evaluating generated code through rapid scanning and application testing, and manual editing. Prompts in vibe coding blend vague, high-level directives with detailed technical specifications. Debugging remains a hybrid process combining AI assistance with manual practices.

Critically, vibe coding does not eliminate the need for programming expertise but rather redistributes it toward context management, rapid code evaluation, and decisions about when to transition between AI-driven and manual manipulation of code. Trust in AI tools during vibe coding is dynamic and contextual, developed through iterative verification rather than blanket acceptance. Vibe coding is an evolution of AI-assisted programming that represents an early manifestation of ``material disengagement'', wherein practitioners orchestrate code production and manipulation, mediated through AI, while maintaining selective and strategic oversight.

\end{abstract}

\section{Introduction and Background}
\label{sec:introduction}
On February 2, 2025, the influential computer scientist Andrej Karpathy posted to Twitter/X describing the idea of ``vibe coding'' \citep{karpathy2025vibecoding}: \emph{``There's a new kind of coding I call "vibe coding", where you [...] forget that the code even exists. [...] I barely even touch the keyboard. [...] I "Accept All" always, I don't read the diffs anymore. When I get error messages I just copy paste them in with no comment, usually that fixes it. The code grows beyond my usual comprehension, [...] but it's not really coding - I just see stuff, say stuff, run stuff, and copy paste stuff, and it mostly works.''}

This description of a new style of programming, that shall henceforth be referred to as the Karpathy canon, has had significant influence. Beyond sparking discourse, it has spurred the adoption of the term ``vibe coding'' to describe an idealised style of programming involving the use of ``agentic''\footnote{Unfortunately, there is no consensus on the definition of the term ``agentic'' or ``agent'' in this context. It is primarily a commercial term rather than a technical term. We invoke this term to make an explicit connection with that commercial discourse, and do not make ontological claims about ``agents'' as a clearly distinct category of AI tools. For our purposes, the property of interest is that tools described as ``agentic'' can typically take a greater range of programmatic actions on the user's content (such as reading code, making direct edits across multiple files, and executing terminal commands) than previous implementations of AI assistance within software development environments.} software development tools such as Cursor, GitHub Copilot Agent mode, Windsurf, Bolt, and others. From the excerpt above, it would seem that the core ideal of vibe coding is disengagement from directly working with code (authoring, editing, reading, etc.) and instead trusting the agentic tool to perform these operations reliably enough for practical use, based on natural language (preferably dictated) descriptions of the vibe coder's intent.

The concept of an idealised philosophy of programming being seeded by an influential figure is something of a trope in programming practice. Prior examples include literate programming, seeded by \citet{Knuth1992-zp}, or egoless programming, seeded by \citet{Weinberg1971-ky}. Perhaps the most novel aspect of the Karpathy canon is its brevity. While literate and egoless programming are each accompanied by a book-length treatise, the Karpathy canon is a 185-word tweetise.

What often occurs following such philosophical seeding is the collective mobilisation of the community of programmers to make sense of, adapt, and adopt (or resist) this proposed way of working. For some, the Karpathy canon merely provides a convenient label for a practice they were already engaged in. For others, the term is an introduction to a new way of working, but with innumerable unanswered questions that need to be negotiated through personal practice on a case-by-case basis. Is it still vibe coding if I manually edit the code from time to time? Is it still vibe coding if I use the keyboard instead of dictation? Is it vibe coding if I use a non-agentic tool? How do I vibe code effectively? What is the spirit of vibe coding, and how does it differ from its letter?

The manner by which a community, through reflexive practice and discourse, comes to hash out these questions and elaborate the concept of vibe coding is a sociological question, which we will not attempt to address in this paper. We raise this to emphasise that, despite nucleating around the Karpathy canon, vibe coding is an \emph{emerging} phenomenon and its properties are being \emph{continuously negotiated} as of this writing. 

Thus, to isolate vibe coding as an object of study, the only criterion we adopt that connects all the disparate activities that programmers engage in while vibe coding or attempting to vibe code is that the programmer self-describes their activity as vibe coding. This may result in false negatives; we might exclude programmers engaging in an activity that to an external observer would have qualified as vibe coding, but which the programmer themselves does not explicitly identify as vibe coding. But with an awareness and appreciation of the breadth of experimentation and negotiation required for a programming community to align on the boundaries of a concept such as vibe coding, we hesitate to characterise any self-identified vibe coding session as a ``false positive'', as that would entail us, as researchers, applying a norm for what constitutes vibe coding where in fact none exists.

As an emerging programming practice and sociological phenomenon, vibe coding is intrinsically worth exploring as it prompts us yet again to revisit longstanding questions in human factors of programming research, such as the role of expertise in programming, programmer agency and control, and so on, much as it was worth characterising the novel nature of programming with earlier generations of code-generating language model tools such as GitHub Copilot \citep{sarkar2022programmingai, lee2024copilotpredictability, barke2023grounded, vaithilingam2022expectation}.

Moreover, it can be argued that programming practice is the canary in the coal mine, so to speak, for how generative AI affects practice in knowledge work more broadly. Programmers themselves are at the frontiers of developing generative AI tools for knowledge workers, and the unique genesis of modern programming practice, forged as it was from an identity crisis in the nature of programming as craft, science, or engineering \citep{Ensmenger2012-qo} has arguably endowed the everyday practice of programming with an unusually high degree of reflexivity. Arguably, programmers are in the privileged position of being among the first wave of knowledge workers to identify new opportunities for applying generative AI to their practice, and also empowered to build these frontier tools for themselves. In essence, where programming goes, so follows the rest of knowledge work.

If the Karpathy canon is taken at face value, vibe coding is our first glimpse at a knowledge workflow subject to material disengagement. By this we mean that the \emph{material substrate} of programming, i.e. code, is no longer worked by the programmer. The programmer deliberately disengages from working in the material substrate to instead orchestrate its production and editing via an AI producer-mediator. It does not take any great leap of the imagination to suppose that a similar disposition or workflow of material disengagement could be applied to all knowledge work tasks, whether writing reports, or emails, or producing spreadsheets, or presentations. Studying how and to what extent material disengagement affects the practice of programming, in its highly natural and externally valid manifestation of vibe coding, gives us a valuable opportunity to anticipate potentially forthcoming changes in the structure of knowledge work.

Thus motivated, \textbf{we contribute the first empirical analysis of vibe coding} as an emerging programming phenomenon, through a framework analysis of curated think-aloud videos of programmers vibe coding sourced from YouTube and Twitch (Section~\ref{sec:method}). Our analysis provides a picture of how goals and intentions are formed during vibe coding, the vibe coding workflow, the nature of prompting, debugging practices, challenges that arise during vibe coding, what kinds of programmer expertise are deployed and how, and the nature of trust in AI during vibe coding (Section~\ref{sec:results}). We discuss how vibe coding appears to represent an evolution of previous generations of AI-assisted coding, the vibe ``gestalt'', the tradeoffs of material disengagement, and how vibe coders attempt to influence societal attitudes to the use of AI (Section~\ref{sec:discussion}).

Our analysis is preliminary; we study only a small set of videos in detail, and the landscape of tools and practices associated with vibe coding is rapidly evolving. Our findings should therefore be understood as a temporally contingent snapshot of how vibe coding appears to manifest at this specific early juncture. Such a vignette serves both as a foundation for future work on understanding the vibe coding phenomenon, and as a reference point for comparison as the practice evolves further.

\section{Method}
\label{sec:method}

\begin{figure}[!t]
	\centering
	\includegraphics[width=\textwidth]{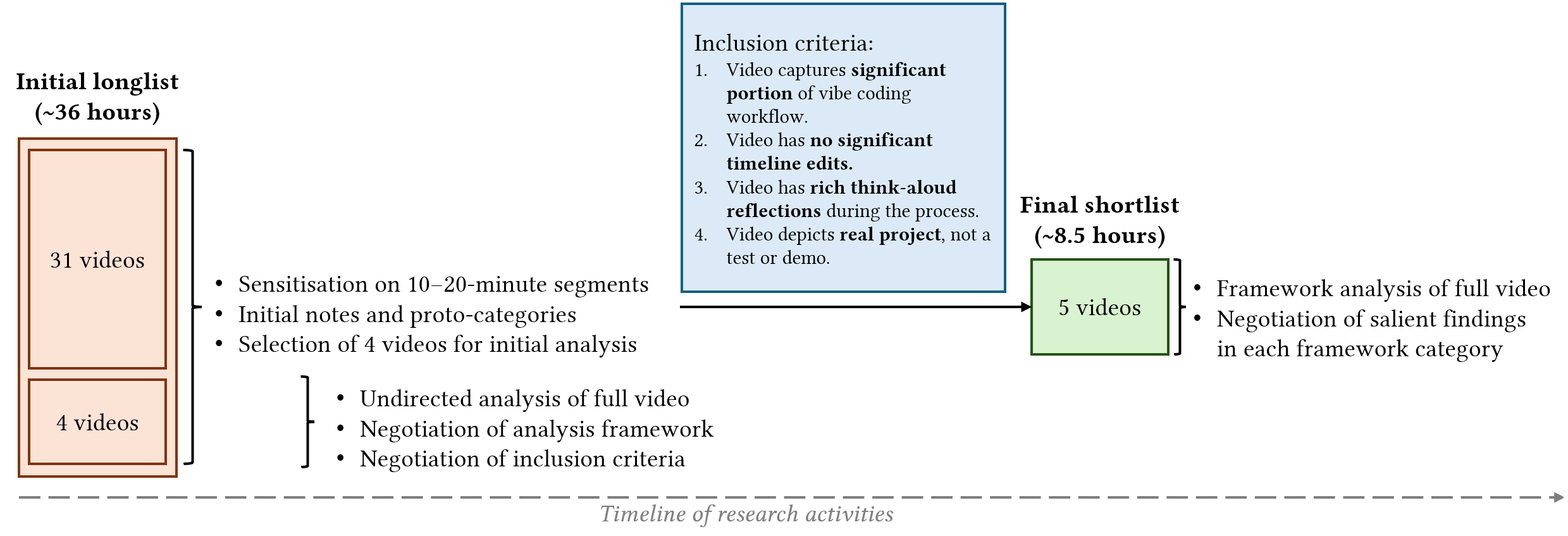}
	\caption{Overview of research process. An initial longlist of 35 videos was analysed in two phases, resulting in a framework for qualitative analysis and a set of inclusion criteria. Applying these criteria, 5 videos (approximately 8.5 hours) were analysed in detail in their entirety according to the framework.}
	\label{fig:study-method-timeline}
\end{figure}

\subsection{Data sources}
Our approach to studying vibe coding through think aloud videos on YouTube and Twitch draws inspiration from prior work. \citet{barik2015heart} demonstrated that online community investigations can ``yield insights into qualitative research topics, with results comparable to and sometimes surpassing traditional qualitative research techniques''. 

Analysis of online sources such as Hacker News, YouTube, Reddit, and other online fora has shed light on aspects of programming practice such as programming as play \citep{barik2017expressions}, the introduction of the lambda function into Excel \citep{sarkar2022lambdas}, and the first generation of code-generating large language models \citep{sarkar2022programmingai}. Just as Hacker News serves as ``an important venue for software developers to exchange ideas as part of a broader cultural ecosystem'' (\citet{barik2015heart}, referring to \citet{wu2014exploring}), YouTube and Twitch function similarly for the vibe coding phenomenon. These platforms are social environments where programming practices are formed and enacted, making them valuable contexts for studying emerging programming practices.

While analysing public discourse regarding vibe coding could help understand how it is \emph{perceived} within programming communities, at this preliminary stage we prioritised studying the \emph{activity} of vibe coding, on the basis that an understanding of how vibe coding is practiced is a necessary prerequisite to understanding any associated discourse. However, with the foundational understanding that this paper provides, an analysis of the discourse would be suitable for future work.

\subsection{Video analysis method}
Given that vibe coding is an emerging phenomenon, we sought the flexibility to document isolated events, sequential event chains, concurrent events, variations in the duration of phenomena of interest, and temporally disconnected but conceptually related events (e.g., when a programmer initiates an activity, transitions to another task, and subsequently resumes the original work). Thus we chose not to segment videos into predefined episodes or fixed time intervals (e.g., 20-second segments, as exemplified in \citet{ragavan2021spreadsheetcomprehension}). Moreover, we required a method that permitted flexible analysis of both verbal think-aloud information and observed actions within the videos.

Consequently, we adopted framework analysis \citep{ritchie2002qualitative,goldsmith2021using}, a form of directed content analysis wherein researchers first familiarise themselves with the data, develop an analytical framework, and subsequently index data according to this framework.

Unlike codes in a qualitative codebook, framework categories do not represent concrete phenomena but rather constitute conceptual domains into which observations can be classified.  This approach thus requires less initial commitment to an ontology of findings. Rather, the indexing process serves to reorganise the data (they are used merely to collate data \emph{about} a concept of interest, rather than data \emph{in support of} a phenomenal theme) so that the directed research questions implied by the framework categories can then be addressed through various interpretive methods. In our case, because the size of recorded observations in each category was relatively small (around 2000 words of combined researcher notes per category), two researchers were able to collaboratively negotiate the most salient findings from each category.

\subsection{Video selection}
We selected an initial longlist of videos by searching the YouTube and Twitch platforms using platform native search using the search terms ``vibe coding'' and ``vibe coding session''. We considered but refrained from including a wider array of search terms (e.g., including tool names such as ``Cursor'') because as mentioned in Section~\ref{sec:introduction}, vibe coding is an emerging practice that is being continuously defined and negotiated and our study is partly an exercise in understanding that process; we thus adopt the methodological principle of ordinary language philosophy \citep{laugier2013we} of attending to the term as it is used and enacted in practice, rather than attempting to impose any \emph{a priori} interpretations of vibe coding, as including other keywords would have done.

Searches were conducted during the period 17 March 2025 to 02 April 2025. Prior to inclusion, videos were lightly inspected to exclude any false positive videos that clearly contained no developer reflections about the concept of vibe coding (e.g., advertisements, slideshows with text-to-speech voiceovers, ambient music videos, etc.) and extremely short videos (less than 3 minutes). This resulted in an initial list of 35 videos with a combined duration of 35 hours, 46 minutes and 21 seconds. 

Next, two researchers independently viewed 10-20 minute sections of each video to sensitise and familarise themselves with the content, making initial notes about potential themes and quality criteria. Through negotiation, four videos were selected for the two researchers to view in their entirety to perform a preliminary qualitative analysis.

Next, the two researchers met to discuss their analyses of the four videos with the objectives of a) determining the quality criteria for videos to be included in the final analysis, and b) determining the categories of the framework analysis.

We determined the following four criteria for inclusion. First, we prioritise videos that capture a \emph{significant portion} of the vibe coding workflow, ideally from end to end (starting with a blank project and ending when the project is complete), so that the researchers can understand the full context of the project and any behaviours displayed during vibe coding. Second, we prioritise videos without significant timeline edits, which video creators often make to shorten videos and improve audience engagement, but which can compromise the value of the video as a reflection of the true vibe coding process. Third, we prioritise videos with rich think-aloud reflections from the developer \emph{during} the coding session, aiming to exclude those with post-hoc voiceovers or those with significant pauses and silences where coding activity is occurring but without spoken reflection. Fourth, we prioritise videos in which the project being worked on has a sufficient level of realism, external validity, or seriousness, as opposed to videos of toy projects made purely for demonstration.

The researchers revisited the longlist and annotated each video according to these quality criteria, negotiating with each other if any uncertainties arose. Out of the initial list of 35, only 6 videos met all four criteria unambiguously. In the intervening time, one of these had been withdrawn from the Twitch streaming platform. Consequently, our final list for deeper framework analysis consisted of 5 videos with a combined duration of 8 hours, 27 minutes, and 19 seconds. Of these, two videos, YT22a and YT22b, were two parts of the same vibe coding project by the same developer -- in our results we therefore combine findings from both under the single ID YT22.

The full list of videos considered, including the four videos selected for the framework development and the five videos analysed in detail are given in Appendix~\ref{apx:video-list}.

\subsection{Framework analysis}
For determining the categories of the framework analysis, the researchers independently compiled notes while viewing videos from the entire longlist on potential categories and research questions to address, resulting in a combined list of 22 proto-categories at various levels of granularity. After negotiation, three categories (on programmer goals, the formation of intention, and the deployment of expertise) were further elaborated to accommodate nuances that researchers had observed at this initial stage. Two proto-categories were merged into a broader category on workflows and tools, and one proto-category was removed, namely, how programmers and non-programmers differ in their approach to vibe coding, because our dataset did not contain any vibe coding sessions featuring a user with no programming expertise. 

Our final framework consisted of 9 top-level categories organised into 20 subcategories. The 9 top-level framework categories were: goals, intentions, workflow, prompting, debugging, challenges, expertise, trust, and definition and performance. An overview of these categories and some of the questions associated with each is given in Table~\ref{tab:framework-overview}. The full framework, including subcategories and elaborations of specific questions, is given in Appendix~\ref{apx:framework}.

Finally, two researchers divided the dataset into two parts of similar duration, and independently applied the framework to qualitatively analyse one part each. In a slight deviation from the standard method of framework analysis where units of the analysed data are directly attached (``indexed'') to framework categories, in this case the researchers made qualitative notes and observations, including excerpts from the think aloud transcripts of the videos, under each category as appropriate. This deviation is necessary to accommodate the difficulties with unitising videos to properly account for emerging phenomena (as mentioned earlier) and also to account for the fact that a phenomenon may be observed by the researcher as manifesting through an irreducible combination of onscreen actions and think-aloud reflection (which may not occur simultaneously). An overview of the research process is given in Figure~\ref{fig:study-method-timeline}. The results of this analysis are presented in Section~\ref{sec:results}.

\section{Results}
\label{sec:results}

\subsection{Goals}
\label{sec:results-goals}
\subsubsection{What kinds of applications are built using vibe coding?}
Our examples spanned varying domains and complexity levels. In the simplest case (YT15), a creator developed an animated diagram explaining a coding agent's workflow. This project focused primarily on visual design elements, including arrow positioning, labelling, animation, and aesthetic styling with specific colour themes. In TW1, the creator created and successfully deployed various member-exclusive features to an existing gaming website. The YT21 creator built a code-base chat application that integrated with GitHub repositories, featuring user authentication, persistent chat histories across sessions, and URL-based navigation to specific conversations. The project applied retrieval augmented generation to answer questions about a user's codebase. In YT22, a developer created a web application to help international students with job applications, incorporating resume tailoring based on job descriptions, user account management, payment processing via Stripe, and file upload functionality.

\subsubsection{Do programmers expect complete or partial success with vibe coding, and to what extent is vibe coding exploratory?}

Users demonstrated a range of expectations when engaging in vibe coding, from full success to partial achievement to exploratory \citep{kery2017exploring} use, sometimes with these expectations evolving throughout their process.

In YT15, the user begins with an expectation of full success, seeking to create a diagram because they don't feel they have the skills to make this diagram themselves. The user ultimately adjusts to accepting partial success, expressing satisfaction with reaching ``80\% of the way,''. They also show an emergent exploratory element, as when the project is completed, they express curiosity about the generated code ("how did we do this?") and express the intent to read the generated code to learn from it.

TW1 also expected full success for vibe coding new features to add to their deployed website. Rather than applying vibe coding in an exploratory manner, their approach combined vibe coding to generate code for each feature with manual fixes and programming to accomplish a set of tasks they had collected on a to-do list.

In YT21, the programmer demonstrates strong expectations of full success. They note that they had already made rapid initial progress before recording, reaching ``80 or 90\% of the way to a working version.'' Their stated intention to incorporate the application into a larger project and eventually productise it for public use all signalled expectations of complete or near-complete success. While the project being ``greenfield'' (i.e., the codebase was built from scratch) allowed ``more of an opportunity to explore,'' this exploration occurs within the framework of building a finished application rather than just experimenting with the technology.

YT22 shows a measured expectation of partial success, specifically aiming for a Minimum Viable Product (MVP). The creator discusses monetisation plans, indicating serious intentions beyond exploration. However, they explicitly state that the outcome doesn't need to be ``production ready''.

\subsection{Intentions}
\label{sec:results-intentions}
\subsubsection{How is the initial intention formed for a vibe coding project?}
In most cases, creators formulated objectives \emph{before} engaging with AI tools, though the specificity and scope of these prior intentions varied. The creator in YT15 approached the session with the goal to generate a diagram explaining a code agent's workflow, having already conceptualised the diagram's structure and components. The programmer in YT21 began with a defined intention to enhance an existing application by adding support for persistent conversations across multiple sessions, having already decomposed this goal into manageable steps. We observed detailed pre-planning in YT22, where the developer had established not only the core functionality but also the technical stack, authentication method, and monetisation strategy before starting to code.

\subsubsection{How and why is intention refined over the course of the vibe coding session?}
\label{sec:intention-refined}

The iterative, conversational nature of vibe coding supports the refinement of intention through progressive discovery and refinement. Across all videos, the creator's ideas expanded beyond their initial vision as refinements were driven by factors such as dissatisfaction with the initial output and emerging needs or ideas during the process of evaluating the output. In the vibe coding workflow, users must simultaneously form intentions for the artefact they are developing and form intentions for how to communicate those goals effectively. The inevitable mismatch between the user's internal intent and the AI ``interpretation'' of it becomes a significant driver for refining not only the project's direction but also the user's prompting strategies and their understanding of AI capabilities.

AI contributes to shaping final by filling implementation gaps and suggesting new possibilities. In YT15, the intent clearly evolved through multiple iterations, starting with an initial diagram layout, developing the placement of arrows, then adding styling and animation. The direction of the project was significantly shaped by the back-and-forth with the system as the generated code variously: succeeded in matching the creator's intent directly, thus allowing progress; or succeeded indirectly such as by interpreting vague instructions, thus sparking ideas; or failed, thus necessitating debugging or a goal ``pivot'' \citep{RubberDuck2024}.

In one instance for TW1, the AI generated code had an error, but on inspection they realised that there was a bug in the previous code that was revealed by the new code. In another instance, the code generated did not match their original intent, but gave them ideas for how it might be useful for implementing future alternatives. 

In YT22, we see a negotiated process where the creator actively evaluated AI suggestions, removing some AI-generated features, like the completion rate box, accepting others, while identifying missing functionalities such as a login button. Similarly, in YT21 testing the application revealed unexpected needs, such as discovering the necessity of a stop button during testing, which arose directly from experiencing the application's behaviour and identifying a usability issue that was not part of the initial plan.

AI tools are also consulted outside of the vibe coding process for guidance and iteration on intent, such as soliciting feature ideas (TW1). Other traditional sources of intention refinement, such as viewing prior art and alternative commercial products as design references, were also observed (YT21).

\paragraph{The blurring of prototyping and production.}
\label{sec:prototyping-production}
Vibe coding seems to significantly blur the lines between prototyping and the initial stages of production. Rather than sketch out ideas in text, diagrams, and other traditional low-fidelity media, it is possible to generate functional code very early in the ideation process. The speed at which a basic, runnable application is established suggests that vibe coding might collapse the boundaries between these traditional development phases. This raises several questions, including to what extent the initial ``vibe'' may set the trajectory for the entire project in a more profound way than traditional prototyping.

\paragraph{Context momentum.}
\label{sec:context-momentum}
Within vibe coding sessions, the history of interactions and the outputs previously generated significantly influence the subsequent direction of code generation. We term this phenomenon \emph{context momentum}. This creates a form of path dependence, where early prompts and the outcomes they yield can steer the project onto a particular trajectory that may become challenging to diverge from later in the session (which is reminiscent of the cognitive dimension of premature commitment \citep{green1989cognitive}, albeit here it is difficult to characterise this as a property of a notation). Context momentum is the cumulative effect of the model's ``interpretations'' and the programmer's responses building upon one another to shape the evolving intention for the program. 

One example comes from YT18,\footnote{This video was not part of our final framework analysis, but this example is given here for its clarity.} a project focused on building an exchange rate application, where the user initially requested historical exchange rate data. The model interpreted this request by providing a date picker to retrieve data for a single date. Even though the user's underlying intention was for a feature that would retrieve a range of dates (which was not articulated explicitly in the prompt), the user found the response satisfactory and proceeded. The model's interpretation thus became the established path. When the user later requested a feature to query a date range (this time explicitly) in a different part of the application, the newly generated code, likely influenced by the prior single-date implementation, still fetched data only for a single date. The user believed this was a downstream consequence of the earlier implementation choice, explicitly linking the current difficulty to the prior model interpretation and its resulting code. Thus, while programmers may choose to accept edits that do not entirely align with their intent, but which can still satisfy their requirements, to maintain speed of development, this may result in greater downstream challenges in communicating intent and steering AI outcomes.

Context momentum may also have positive, exploratory outcomes. For instance, YT15's decision to add animations was an impromptu goal that appeared to arise on the basis of the possibilities presented by the existing structure in the generated code. The possibility of the developer formulating such a goal thus appeared to be contingent on a certain set of arbitrary but satisfactory implementation decisions made by the model, and a different set of satisfactory decisions may have led to different improvised goals.

\subsection{Workflow}
\label{sec:results-workflow}
\subsubsection{What are the stages of the vibe coding workflow?}
The vibe coding workflow consistently begins with an initial goal-setting and prompting phase. In YT15, this involved the creator giving a detailed spoken prompt to Cursor via a transcription tool to generate the initial diagram, establishing a diagram explaining the code agent's workflow as the first goal. Similarly, the creator in YT21 started with a clear goal: to modify his existing `chat with a code base' application to support multiple chat conversations per repository, which they subsequently decomposed into smaller sub-goals. In YT22, the creator sketched a flow diagram to visualise the tool's functionality and applied Code Guide,\footnote{Available at: \url{https://www.codeguide.dev/}. Last accessed: 19 September 2025.} a tool that generated a project plan and documentation from his app requirements.

Following initial generation, developers consistently engage in evaluation and iterative refinement cycles. In YT15, after quickly previewing the diagram, the creator notes that it ``meets requirements'' in terms of basic functionality but ``is not very pretty,'' prompting refinement of specific elements (e.g.,``arrows are better, but could be even better''). Throughout TW1, the creator reviewed the model's code changes, and upon discovering any issues prompted the model about what was incorrect and regenerated code. YT22 similarly shows the developer's continuous review of the AI-generated code with manual editing to fix errors, remove unwanted features and adjust functionality. This evaluation phase frequently involved test running the application, such as frequently switching to the browser to test the implemented changes and verify if they met expectations (YT21).

Thus, vibe coding is characterised by \emph{iterative goal satisfaction} \citep{RubberDuck2024} cycles, where developers (1) Formulate a goal or sub-goal, (2) Prompt the model to generate code to achieve that goal, (3) Review the generated code, (4) Accept or reject the changes, (5) Test the changes in the application, (6) Identify any bugs or areas for improvement, (7) Refine the prompt or transition to manual debugging and editing. Steps (2)-(7) are repeated until the sub-goal is satisfied or pivoted, at which point the programmer returns to step (1). The entire process is repeated until the programmer decides not to formulate any further goals.

\subsubsection{Portfolio of tools and technologies}
An ecosystem of interconnected tools is used with AI-integrated development environments at the centre. In YT15, Cursor serves as the primary vibe coding platform, providing an interface for prompting, code generation, and editing, allowing users to switch between models such as o3-mini and Claude 3.5 Sonnet. The creator also uses a transcription tool to convert voice instructions into prompts for Cursor. A web browser is used for previewing and evaluating the generated HTML and JavaScript code.

YT21 demonstrates a similar pattern. The programmer uses Cursor for code generation and editing, switching to a web browser to visualise and test the web application. This programmer actively used browser developer tools (inspector, console, network tab) to identify error messages and check API calls. Additionally, the terminal is used to monitor server-side processes and error messages.

The YT22 creator begins with Excalidraw for initial planning and conceptualization, creating flow diagrams of the intended tool. This is followed by using Code Guide, a meta-level tool that generates project plans and documentation intended to reduce AI hallucinations by pre-processing prompts to introduce structure and detail. Only then does the creator move to Cursor for code generation in agent mode, intentionally configured with specific models like OpenAI's GPT-4o. The terminal is used for project setup and running development servers, while the web browser serves for testing, verification, and accessing documentation.

\subsection{Prompting}
\label{sec:results-prompting}

\subsubsection{How do creators set up a vibe coding session for effective prompting?}

To help their prompts be more effective, creators modify the system instructions for the code-generating language model, compare different IDEs to evaluate the fitness-for-purpose of their AI integration, compare different code-generating models for their speed, quality, and cost-per-token, and pre-process prompts to reduce hallucinations.

The creator in TW1 takes a formalised approach to prompt customisation, starting with an existing template called \emph{.cursorrules} sourced from GitHub. They incorporate this into their own 
\emph{Claude.md} file, which serves as system instructions for the code-generating model. They personalise these instructions by specifying the programming languages, frameworks, and libraries used in their codebase.

YT21 reveals more of the implicit preparation required for effective prompting during vibe coding. The programmer selected Cursor specifically for its AI integration after evaluating alternatives like VS Code and Replit. The programmer had already mentally mapped their goal of making the chat interface ``more like ChatGPT'' and broken it down into smaller steps, starting with ``creating an API for a new conversation.'' Their familiarity with the codebase and pre-selection of Claude 3.5 Sonnet as their model represent forms of preparation, even if not explicitly framed as setup work.

In YT22, the creator uses Code Guide to input requirements and generate a project plan and documentation, which they review and refine by updating version information. This preparation is strategically intended to help the model hallucinate less. They actively switch between Claude 3.7 Sonnet and OpenAI GPT-4o based on considerations like output speed and token costs (explicit experimentation and comparisons of performance were also observed in YT15 and YT21).

Vibe coding does not appear to be dependent upon, or associated with, any particular level of prior development of the codebase. Developers in our dataset variously started vibe coding entirely from scratch starting with an empty directory (YT15), or from a boilerplate starter template (YT22), or \emph{in medias res} for continuing to develop a pre-existing codebase (TW1, YT21).

\subsubsection{What explicit prompting strategies are used during vibe coding?}

We observed several explicit prompting strategies, such as management of the context, referring to specific code elements by name, structuring instructions into lists, outcomes, and constraints, pasting in examples and error messages, referencing previous output, pre-processing or generating prompts through external tools, deliberately limiting the scope of prompts, and including external documentation as resources.

We observed deliberate management of the AI's working context. When shifting focus to animating arrows, YT15 opened a new ``composer'' window (essentially, a new chat thread within Cursor) to create a fresh context. Similarly, Before starting new phases, YT22 closed all tabs to ``clear the context from the AI''.

The YT21 developer frequently provides context through code snippets and refers to specific code elements by name, using names of functions, variables, and APIs to direct the model with precision and granularity. They also structure complex instructions using numbered lists and explicitly state desired outcomes and constraints. They also provide concrete examples, e.g., pasting an example of the expected JSON response from an API to guide implementation. This developer also explicitly referenced the model's previous output in new prompts to provide corrective feedback. Uniquely in our dataset, this developer employed inline prompts for localised changes directly within the code editor (i.e., the smaller-scoped, autocompletion-style of interaction typical of earlier generations of coding assistants) for targeted modifications.

As mentioned, the developer in YT22 began with documentation and implementation plans generated by Code Guide as context for prompting Cursor. TW1 also demonstrates a meta-prompting approach by using one model to generate a prompt for the coding AI agent. The developer in YT22 also adapts community knowledge, using a prompt structure found on Twitter but tailoring it to their specific project context. They explicitly practice scope limitation, focusing on one ``phase'' of the implementation plan at a time to reduce hallucinations. In another form of scope limitation, they also introduce constraints to prompts such as ``don't integrate stripe yet. Just make a design with dummy data'', perhaps anticipating and preempting the model's tendency to fill in unstated gaps, or tendency to set an overambitious goal that the model is likely to fail to achieve in a single step.

The developer in YT22 demonstrates further fine-grained guidance methods, particularly by providing external documentation as context, explicitly copying the URL of the official Tanstack Form documentation with instructions to ``use docs,'' employing the model's ability to index external resources. When encountering runtime issues, they supplement prompts with error messages, directly pasting browser errors and asking for fixes.

\subsubsection{What is the granularity of prompts and iterations? Do users address a single objective at a time with prompts, or issue multi-objective prompts?}

The granularity of prompts in vibe coding spans a spectrum from high-level directives to extremely detailed, low-level instructions, spanning from single-objective prompts, to complex, mixed-objective prompts where different unrelated requirements are expressed simultaneously and at different levels of detail.

In YT15, the developer begins with detailed spoken instruction to establish initial requirements for a diagram, including modules and data flow specifications. When results don't meet expectations, they employ iterative refinement with precise feedback. When trying to prompt the model to produce an animation, they provided a detailed step-by-step description specifying the exact sequence of visual elements they expected to see. 

The creator in YT15 uses both precise and imprecise language to convey aesthetics. When requesting arrow repositioning, the creator gives specific directions like ``not to use diagonal ones'' and that ``arrowheads should be pointing to the middle of the boxes.'' However, the creator also employs imprecise descriptors such as expanding space ``significantly,'' relying on the model's interpretation rather than providing exact measurements. The creator also issues a high-level prompt mentioning a desired theme and colour palette without specifying details, leaving these to the interpretation of the model. Similarly, the programmer in YT22 uses a mix of specific and ambiguous instructions for styling, such as ``move the date below and use a very small font'' and ``don't use bold font, narrow the space between conversations.'' These prompts combine specific actions with somewhat subjective descriptors like ``very small'' and ``narrow,'' leaving some interpretation to the model while maintaining directional control.

YT21 shows a similar pattern but with explicit reflection on the strategy. The programmer initially uses very brief, high-level prompts such as simply requesting to ``create an API to create conversation.'' When these broad prompts fail to yield the desired results, the programmer acknowledges, ``I should have been more clear here,'' and shifts to specific, low-level prompts that target particular functions, variables, and styling elements. The programmer also employs extremely fine-grained iterations for localised changes, such as using inline prompts to modify specific parts of functions. The programmer's statement that they ``start by asking very dumb questions. And if it gets confused, [they] start providing more detail'' reveals an intentional approach to granularity adjustment.

The developer in TW1 attempts to keep prompting succinct yet detailed, but ultimately finds longer prompts to be better so as not to give the model ``a lot of leeway to go off the rails.'' This developer strikes a balance between high-level directives (e.g, ``Increase the exp for VIP members to 1.3x on the back-end'') while providing specific implementation details when necessary (e.g., which files to edit, and specific callouts about variables or code types within these broader requests).

YT22 intentionally modifies high-level prompts to focus on smaller chunks (``just do phase one'') to reduce hallucinations, revealing a mental model of how prompt granularity affects AI performance. Their debugging prompts were consistently single-objective (pasting specific errors with requests to fix), and even feature requests focused on particular elements like ``Make a page with all resumes that run through the optimization'' rather than requesting multiple features simultaneously. YT22 also reveals situations where extremely fine-grained prompting may be deemed inefficient; the creator often makes manual edits rather than prompting for minor adjustments.

\subsubsection{What is the role of different input modes such as voice, text, and images, in vibe coding prompting?}

Despite the emphasis on voice input in the Karpathy canon, typed text is the primary mode of prompting, but we do observe instances where transcribed speech and images are used as well. Typed text itself may also contain a mix of elements, such as instructions, copied and pasted error messages, and links to documentation (as previously described).

In YT15, the developer primarily uses voice prompting via a transcription tool, though they note the ingrained habit of typing prompts, and consciously try to stay in voice modality, citing the desire to adhere to the Karpathy canon. Despite this, they still switched to text input at times. They attempt to use a screenshot to better communicate the desired changes, though this was unsuccessful due to limitations of the model they had selected at the time (we did observe successful use of screenshots in other videos in our extended corpus). 

In contrast, YT21 and YT22 dominantly use typed text. YT21's creator relied virtually exclusively on text-based prompts within Cursor's composer window and inline chat feature, with no evidence of voice or image inputs. Similarly, in YT22, text is the primary mode of input with the creator typing new prompts and modifying existing ones through typing.

Code reuse strategies appear consistently across the examined videos, naturally through textual means. In YT21, the creator frequently copied and pasted entire function bodies into the composer window to provide context for their natural language instructions. Similarly, the creator would copy error messages from the terminal or browser console into the prompt window. The inclusion of expected output formats also appears as a code reuse strategy in YT21, where the creator provided an example of the expected JSON response to guide the model.

Notably absent from these observations are examples of creators directly borrowing code from online sources like Stack Overflow or GitHub to guide the model, though the practice of sharing documentation URLs in YT22 suggests that external resources do play a role.

\subsection{Debugging}
\label{sec:debugging}
\subsubsection{What debugging strategies are applied during vibe coding?}
Across the observed vibe coding sessions, developers employ a blend of traditional debugging techniques and AI-assisted approaches. The debugging process typically begins with visual inspection of the code and test running the application. This initial assessment is followed by more detailed debugging strategies depending on the nature of the issues encountered.

AI ``hallucinations'' and failures to fully adhere to prompt instructions can create errors. In YT21, the model sometimes generated code with non-existent properties, requiring the developer to identify these issues through error messages and code inspection. The developer in TW1 expressed frustration when generated code did not adhere to the system instructions they defined, noting that it exported a function when it didn't need to. In YT22, the model produced documentation referencing Next.js 14 and JavaScript files when the creator was using Next.js 15 with TypeScript, requiring manual updates.

In YT15, the creator primarily relied on identifying visual discrepancies between the desired outcome and the generated output. For example, they noticed issues such as unsatisfactory arrows in diagrams, a canvas that was too small, and overlapping arrow labels. This approach is particularly relevant for graphical or UI-focused projects, where the correctness of the code is largely judged by its visual output rather than its internal logic.

More technically sophisticated debugging approaches appear in YT21 and YT22, where the programmers heavily utilised browser developer tools. In YT21, the programmer examined error messages in the browser's console and network tabs to pinpoint issues and verify API calls. Similarly, in YT22, when encountering a blank screen, the creator immediately opened the browser's developer console, identifying a ``trpc error''. Terminal analysis was also observed in YT21, with the programmer monitoring terminal output alongside browser tools.

A distinctive feature of vibe coding is the use of the AI itself as a debugging tool, or perhaps more accurately, to bypass the manual debugging process entirely. In YT22, the creator copied error messages directly from the browser console and pasted them into the Cursor chat prefixed with prompts like ``Please fix it'' or ``Refer the docs to fix this error.'' Similarly, in TW1, the creator asked the AI to ``find and debug the issue and fix it.''

Creators frequently employ iterative refinement through targeted prompting. In YT15, the creator provided detailed instructions for repositioning arrows, fixing overlapping labels, and adjusting animations. This strategy of using natural language prompts to guide debugging is also observed in YT22, where instead of directly modifying code, the creator issued prompts such as ``I don't need the function of pasting the URL of job description. Please use the text approach only.'' TW1 employed a strategy of resubmitting a prompt after iteration on the system instructions to guide the model to produce results in line with their preferences.

Creators still engage in manual code review and editing. In YT21, the programmer frequently scanned and read the AI-generated code to assess alignment with expectations and identify potential issues before testing. Similarly, in YT22, the creator frequently examines the generated code, scrolling through files and looking at diffs to understand the logic of the AI output. In TW1, the creator inspected the AI-generated code to check if it was viable, and in some cases if the output was ``close'', they accepted the code with the intention to fix it manually.

Users often exhibited mixed manual-automated debugging strategies involving hypothesis formation and strategic prompting. In YT21, when errors occurred, the programmer formulated hypotheses about the cause based on error messages and observed behaviour, then verified these hypotheses through browser or code inspection. This was followed by specific and targeted prompts containing instructions on how to fix identified issues, such as prompting the model to ``pick the repo\_ID from the URL''.

Model switching is occasionally used as a debugging strategy. For instance in YT15 the creator switched from the o3-mini model to Claude 3.5 sonnet when trying to animate arrows. This suggests that creators develop mental models of the relative capabilities and proficiencies of different language models.

Users sometimes acknowledge and deliberately ignore issues, as in YT15, where the creator explicitly chose not to address a compiler warning (``do not use empty rule sets'').

\paragraph{The special nature of code reading in vibe coding.}
\label{sec:debugging-reading}

A striking aspect of how programmers engage with AI-generated code in vibe coding, particularly during debugging, is the speed and high-level nature of their code inspection. Rather than line-by-line code review, we observed programmers rapidly moving from point to point, performing what could be described as impressionistic scanning rather than a linear read. This allows for a rapid, almost immediate assessment of whether the generated code ``looks all right'', ``meets requirements'', or is ``exactly what I asked for'' (YT15).

This speed is facilitated by several key techniques. Programmers pay close attention to the visual code diffs presented by the IDE (e.g., Cursor), which highlight additions in green and deletions in red. They can quickly assess the volume of changes by looking at the size of the red and green highlights and make a decision to accept changes without too much careful scrutiny if the overall shape and size of the changes seem acceptable, or reject them if not. In one instance, the programmer accepted a large number of diffs within two seconds because they could immediately tell it was what they wanted (YT21). 

Programmers are drawn to specific function calls and the code comments written by the LLM (e.g., YT21). They look for key indicators of task success, such making sure that the right API is being called or the right identifiers are being used (YT21), inspecting lines of code in more detail if something catches their eye during this rapid scan. Especially with structured code like HTML or React components, programmers can quickly assess the overall structure. 

Experienced programmers can look at AI-generated code and intuitively sense if it's using the right level of abstraction. Observing the AI writing low-level HTML/JS/CSS primitives for diagrams, the programmer immediately recognised the pattern and felt like they were ``reinventing the wheel'', speculating that there ``should be some libraries that do these things'' (YT15).

The ability to see beyond the immediate AI suggestion and understand its implications across the project is a critical use of expertise. Expertise allows programmers to cross-reference AI-generated code with other parts of the codebase or external documentation. They might check a specific file to ensure the AI used the correct API path (YT21) or cross-reference parameter names used by the AI against functions in different files (YT21, YT15).

The expert ability to mentally visualise the finished product \emph{through} the code without needing to test run the application also supports the rapid scanning of AI generated code for verification. For example, YT21 repeatedly glanced at HTML to ensure generated code appeared in the right section, immediately spotting where the data is going to be displayed or where the title is going to be displayed simply by looking at the code structure.

Moreover, the process of rapid scanning serves for more than simply allowing programmers to react to potential errors in the current round of AI-generated edits. In reading AI-generated code, programmers also exhibit a \emph{proactive stance}. Programmers consciously prepare themselves for reviewing subsequent cycles of AI output. By familiarising themselves with the overall structure of generated code as it is generated, noting key identifiers, components, and implementation decisions made by the model, they prepare themselves to assess \emph{future} outputs and diffs, maintaining their ``glanceability'' as the codebase grows. They might reference API documentation or component structures to know what the model \emph{should} produce in certain areas (YT21, YT22). 

A related proactive behaviour is the programmer's ability to hypothesise about situations such as needing to make changes to an additional related file, even though Cursor hasn't suggested it directly (YT21). This requires an understanding of the codebase's interconnectedness and dependencies that are not direct semantic dependencies, but rather contained in the programmer's mental schema, and therefore invisible to the code-generating model.

\subsection{Challenges}
\label{sec:results-challenges}
\subsubsection{What challenges besides code bugs do programmers encounter while vibe coding?}

Practitioners face several challenges when engaging in vibe coding beyond traditional code bugs. These challenges span technical, conceptual, and workflow dimensions.

One challenge is communicating visual and abstract ideas. In YT15, the creator struggled with articulating what they wanted to do with the arrows, encountering the fundamental difficulty of trying to talk about diagrams using words. Complex conceptual goals can also be difficult to articulate as prompts, as seen in TW1 where the developer faced difficulties forming the prompt to describe a matchmaking feature, asking ``how is [AI] going to give me what I want if I don't even know?''

Understanding AI capabilities and limitations is another challenge. In YT15, the creator expressed uncertainty about whether a specific AI model (Claude 3.5 Sonnet) could create an animation as desired and later switched to a different model (o3 mini) after an unsatisfactory experience. In TW1, the creator felt that some models were better at certain workflows, indicating that matching the right AI capability to the right task requires experience and knowledge with the models. TW1 also stated that ``LLMs will get you most of the way there, but they don't get you all the way there'', perceiving universally inherent limitations in AI capabilities. YT22's need to include specific documentation for Tanstack Form to ensure proper integration also demonstrated that the model's ``general knowledge'' isn't always sufficient for specialised frameworks.

Tool-specific limitations also present challenges. In YT15, the creator didn't know how to revert changes made by Claude and struggled to navigate Cursor's interface to find a checkpoint. In YT21, the programmer faced a situation where Cursor displayed generated code changes as a chat message (rather than edits in the file), making integration difficult.

Deciding when to rely on AI versus manual intervention (discussed further in Section~\ref{sec:transitions}) introduces new metacognitive load \citep{tankelevitch2024GenAImetacognition}. In YT21, the programmer frequently reviewed and sometimes rejected AI-generated code, stating that reviewing changes helped them remain in control. In YT22, the creator frequently switched between prompting and making manual edits without clear criteria for choosing one approach over the other. The developer in TW1 observed that for small adjustments, they felt faster working manually, but for more substantial edits, they preferred using AI assistance.

\subsubsection{How are these challenges addressed?}

Iterative refinement of prompts is almost inevitably the programmers' first recourse when encountering communication challenges with the model. In YT15, when the creator struggled to articulate visual intent for arrow placement, they iterated through multiple prompts with increasingly detailed spoken instructions. Similarly, in YT21, the programmer recognised when initial prompts were too vague and provided more detailed and specific prompts in subsequent attempts, often including numbered lists of requirements and referencing existing code snippets. When the model ``did not listen to the system instructions'', TW1 refined the system instructions to emphasise acceptable behaviour; a higher-order strategy than refining individual prompts.

At one point, the programmer in TW1 adopted an unusual prompting strategy where the developer responded to an unsatisfactory result by prompting ``Bro, you are losing aura with this. Come on.'' before pasting the original prompt and resubmitting. This explicit ``vibification'' produced more code than the previous turn, though the programmer in TW1 still expressed dissatisfaction with the code quality. This suggests that some vibe coders experiment with social or emotional appeals, albeit perhaps tongue-in-cheek, in prompts. We also observed similar styles of prompting in other videos in our broader corpus. The results of such appeals, as might be expected, are mixed.

Developers strategically switch between different AI models based on their perceived strengths. In YT15, the creator actively switched between OpenAI's o3-mini and Anthropic's Claude 3.5 Sonnet models based on their qualitative assessment and the perceived strengths of each for different aspects of the task, using one for animation and another for initial code generation. Similarly, in TW1, the developer discussed where they used models from the Claude family to generate code for a web page, but switched to OpenAI's GPT 4.5 to rewrite the text because they believed its text quality was an improvement over Claude.

\subsection{Expertise}
\label{sec:results-expertise}
\subsubsection{When and in what ways do experts deploy their expertise in vibe coding?}
Expertise is consistently deployed throughout the entire vibe coding process, though in ways that differ from traditional programming.

In the initial phases of projects, experts apply their knowledge to select appropriate tools and models. In YT15, the creator demonstrates expertise when evaluating different AI models based on their capabilities for diagram creation. Similarly, in YT21, the programmer makes a deliberate choice to use Cursor over the Visual Studio Code or Replit IDEs based on an expert assessment of their affordances.

During code generation and review, expertise manifests in rapid evaluation and error detection. In YT15 the creator applies their expertise to rapidly assess HTML code visually. Similarly, in YT21, the programmer consistently reviewed the code for key elements and potential issues, making immediate judgments about the correctness and suitability of the suggested code (as described in Section~\ref{sec:debugging-reading}), demonstrating a high level of familiarity with programming languages and coding patterns.

Expert knowledge also guides problem identification and solution development. In YT22, upon encountering a blank screen, the creator quickly inspects the browser console and identifies a `trpc error,' demonstrating expertise in reading and interpreting error messages. In YT21, the programmer displays expert knowledge of debugging techniques by immediately recognising database issues from error messages, effectively using browser developer tools to diagnose problems, forming hypotheses about root causes, and inserting print statements to trace code execution. YT21 also demonstrates understanding of API endpoints, data structures, and how different parts of the application communicate, allowing them to identify issues with API calls. Similarly, in YT22, the creator identifies that an application is trying to connect to PostgreSQL instead of Supabase and knows exactly how to manually edit environment files to correct the configuration. In YT22, the creator evaluates AI-generated code against expected standards by checking generated React components against documentation, demonstrating and applying their knowledge of UI libraries.

Expertise guides quality control and feature alignment. In YT22, the creator recognizes and decides to remove unwanted AI-generated features that, while not buggy or intrusive, don't align with the intended design. In TW1, the creator demonstrates understanding of when generated code might incur technical debt, which ``lowers developer velocity'', and addresses these issues proactively.

Finally, experts also deploy their knowledge to determine when to transition between AI guidance and manual editing (discussed in more detail shortly in Section~\ref{sec:transitions}).

Thus we observe that expertise is not replaced but rather redirected: from writing code directly to evaluating, guiding, and refining AI-generated solutions. The expert assumes more the role of director, reviewer, and editor than a line-by-line author, but their technical knowledge remains essential throughout the entire development process.

\subsubsection{What kinds of expertise are used in vibe coding?}

As previously mentioned, all the videos we studied featured programmers clearly in possession of expertise in software development. YT21 leverages a range of expertise including specific technical knowledge of databases and programming languages, but equally important are their debugging skills, understanding of API calls, and software architecture concepts. Similarly, YT22 displays technical knowledge of their stack (Next.js, TypeScript, etc.). They make informed decisions about which libraries to use based on project needs and can identify when AI suggestions align with or deviate from requirements.

Beyond ``traditional'' programming expertise (code quality and maintainability, fault localisation and debugging), the programmers we studied exhibited expertise in two additional domains: AI expertise (understanding models, prompting practices, concepts such as context windows, and limitations of AI code generation), and product management expertise (forming and translating goals into features for the product). Perhaps as a consequence of the blending of prototyping and production (as discussed in Section~\ref{sec:prototyping-production}), vibe coding requires (or at least is greatly facilitated by) a broad skillset spanning traditional programming skills, AI literacy, and product vision.

Across these videos, we see creators combining traditional coding knowledge with adaptation to new tools and workflows. Effective vibe coders appear to be those who can fluidly move between writing/editing code directly, formulating effective prompts for AI tools, evaluating AI outputs, and understanding the broader product and user requirements, suggesting both technical foundations and a ``technology-forward'' mindset are valuable components of expertise in this context.

Moreover, we speculate that these programmers are also developing a sense of what might be called \emph{ambient competence}. This is a sense of competence wherein the programmer feels capable of tackling tasks they wouldn't have considered before, because they have access to an AI system that may be able to accomplish the task with very little effort invested, regardless of whether they themselves have the skills to tackle them manually (which they very well may). This can be thought of as the competence that arises as the result of the active fulfilment of the ``awareness of the possible'' (\citeauthor{sarkar2023eupgenai}, \citeyear{sarkar2023eupgenai}; 
 \citeauthor{sarkar2023simplicity}, \citeyear{sarkar2023simplicity}). This could lead to a shifting locus of agency, where the programmer's confidence and the scope and complexity of projects they undertake are increasingly defined by their perceived ability to prompt and guide a model, rather than their own direct coding abilities.

\subsubsection{When do practitioners transition from vibe coding to manual work?}
\label{sec:transitions}

Developers transition from AI-assisted vibe coding to manual work under several distinct circumstances and for several reasons. They might choose to work manually for efficiency, debugging, refinement, and specific episodes where the model is difficult to ``steer''. The transition to manual work during vibe coding is not merely a function of technical necessity but also reflects individual preferences, expertise levels, and expectations for human-AI interaction.

For straightforward edits where the overhead of prompting exceeds the effort of making the change directly, developers often opt for manual work. In YT21, the programmer attempted an inline edit with AI but then decided to perform the simple one-line edit manually after determining that AI assistance was unnecessary for such a minor change. In TW1, the developer articulates their decision-making process about when to work manually versus using AI assistance. In one instance, they switch to manual work when they believed it would be a more efficient workflow, noting that ``IntelliJ's autocomplete and search is still faster than LLMs''. This leads them to ``race the AI'' to find code locations, sometimes stopping AI generation mid-process when they locate what they need faster.

Another common transition point across videos is when AI generates a mostly-complete solution that requires refinement. This can be viewed as a special case of transitioning to manual work for efficiency. For example, TW1 accepted code when the AI ``got close'' to their goal and then manually edited the generated code to be more inline with their preferences for code. TW1 describes their viewpoint that LLMs should be used to ``get close to what you want'' but that expecting perfect outputs is ``a mistake'' -- instead, their optimal workflow involved making post-generation adjustments to generated code, and reuse of generated code in similar features by quickly modifying code to other contexts rather than issuing refined prompts.

Similarly, in YT22, the creator performs manual edits to remove unwanted UI elements, delete redundant buttons, and make small edits in the code to better align with their vision. For instance, the creator manually removes functionality for uploading job descriptions via URL to focus solely on text-based inputs, representing a conscious design decision that required direct code manipulation.

Debugging is as a common trigger for manual intervention across multiple videos. In YT21, the programmer switches to manual work almost immediately when encountering errors, using traditional debugging techniques like browser developer tools, console inspection, and hypothesis formation. In one instance, they hypothesised a solution based on an error message and implemented a fix without AI assistance. Throughout, YT21 frequently engages in manual code review and rejection of AI suggestions based on their understanding of correctness and project requirements.

YT22 shows similar patterns, with the developer manually editing environment files to fix database connection errors and directly intervening to correct runtime issues. YT22 also manually corrects a misalignment in AI-generated documentation (changing "Next.js 14" to "Next.js 15") and deletes redundant steps in implementation plans. These examples suggest that when errors occur, developers often rely on their expertise and traditional debugging approaches rather than AI tools.

However, YT15 presents a contrasting case where the developer states explicitly that they are ``not going to try and fix it on his own'' when encountering an error, citing adherence to the Karpathy canon. The creator in YT15 deliberately avoids manual debugging and relies almost exclusively on re-prompting for changes throughout their workflow. Thus, the videos do exhibit individual differences in approaches to manual work.

\subsection{Trust}
\label{sec:results-trust}
\subsubsection{What is the nature of trust in vibe coding and how is it developed?}
Trust in vibe coding appears to be granular, dynamic, contingent on review, and evolves through interaction with the system, manifesting as a tension between efficiency and comprehension.

In YT15, we see a developer who demonstrates initial trust by accepting AI-generated code with minimal inspection. Trust further develops through subsequent successful outputs. When the model successfully generates complex animations, the creator expresses surprise (``It would feel a little bonkers if this actually works'') followed by reinforcement of trust when it succeeds (``wow, [...] that's exactly what I asked for''). This indicates that dramatic positive outcomes can significantly strengthen trust in AI capabilities.

YT21 presents a practitioner with what appears to be a well-calibrated trust mindset. This developer articulates that they can ``quickly get 80-90\% of the way there with AI''. However, this trust is contingent, as they explicitly state, ``I like to review the changes because it helps me remain in control to some extent.'' The risk of overtrust is actively mitigated as the developer frequently rejects AI suggestions that use non-existent APIs or approaches they considers less maintainable. Frequent (albeit lightweight) code review was observed across all participants. Thus, reviewing changes appears to be important for vibe coding, not only for maintaining understanding of the code, but also for maintaining agency, authorial ownership, and trust.

YT22 further reinforces this notion of contingent trust. The creator explicitly states that they don't believe in ``blindly following the AI'', stating the importance of reading and reviewing the code. They further state: ``AI is just a tool.'' This invocation of the tool-user relationship can be viewed as being in explicit contrast to delegation of responsibility. The developer consistently reviews code, compares it to documentation, and tests frequently, showing that trust is continuously re-established rather than statically assumed.

TW1 illustrates the nature of trust in relation to expertise. Their expertise enables the recognition of ``way too complicated code'' generated by the model and enables them to fix it, while speculating that ``the average vibe coder might not be able to fix that'' without expertise such as TW1 possessed.
This suggests that trust in vibe coding may be significantly influenced by the user's own technical expertise. When discussing a complex codebase of ``150 thousand lines of good code'' the creator emphasises that they cannot allow ``bad'' code to enter the repository. Even on ``simpler'' features, such as an information page that describes the benefits of a VIP membership to the website, TW1 still notes they will review the pull request before committing and deploying to the website. 
Interestingly, this disposition was challenged by a viewer of TW1's twich stream, saying that ``code review is not vibe coding'', which suggests a tension between trust and verification in the vibe coding paradigm.

Users develop trust in AI-generated outputs through a process of experimentation, verification, and adaptation. The risks of overtrust emerge when users fail to critically evaluate AI outputs or lack the expertise to identify problematic code. Across these videos, we see that developers, for the most part, appear to successfully avoid overtrust. They maintain a critical stance toward AI outputs, verify results through testing and review, and possess sufficient expertise to identify problematic code.

\subsection{Defining and performing vibe coding}
\label{sec:results-definition-performance}
\subsubsection{How do programmers define vibe coding?}
Vibe coding thus appears to be a programming approach characterised by conversational, iterative interaction with AI tools, where the model handles significant portions of the coding work. TW1 directly defines it as having ``the AI do the heavy lifting'' and describes the workflow as having the AI creating a feature and then ``spruce it all up.'' In YT21, the programmer describes vibe coding as ``just chatting with the app, saying do this, do that''.

TW1 presents vibe coding as flexible and creative, likening it to ``a Bob Ross painting'' where ``you can do what you want'' when referring to the effort a vibe coder spends on the workflow, suggesting that the defining characteristic may be freedom and creative flow rather than the use of specific tools. For planning activities, YT15 demonstrates that vibe coding can encompass emergent intent where requirements evolve through interaction with the AI. The programmer introduces new requirements like arrow labels and animations after seeing initial results, planning fluidly throughout the process rather than strictly beforehand. This suggests that even when not strictly in an exploratory programming setting \citep{kery2017exploring}, vibe coding is accommodating of an exploratory element.

Based on these observations, vibe coding appears to occupy a higher position on the spectrum of AI reliance in programming activities than ``traditional'' AI-assisted coding (e.g., as with the initial generation of GitHub Copilot as documented in \citet{sarkar2022programmingai} or \citet{barke2023grounded}). It goes beyond using AI for discrete tasks or code completion, involving significant delegation of code creation and modification to AI systems. However, it is not entirely hands-off. As seen in YT21, programmers may rely on AI for code generation but readily switch to manual debugging and problem-solving when necessary. Similarly, YT22 emphasises that human oversight remains important, with the programmer insisting on reading and reviewing the AI-generated code.

The position on this spectrum also appears to vary based on individual preference and project phase. TW1's comment: ``don't review just submit'' (albeit humorously, as TW1's workflow relied heavily on reviewing AI-generated code), suggests some practitioners might push toward greater AI reliance, while others, like those in YT21 and YT22, maintain more balanced approaches. YT15 shows that programmers might be more willing to accept AI-generated code quickly in early stages but become more critical as development progresses. YT22 initially describes vibe coding as when ``you don't really dig into the code, you just embrace the AI,'' suggesting minimal human intervention. However, the same programmer later clarifies that they ``don't believe in blindly following the AI'' and emphasises the importance of reading and reviewing code. This contradiction reflects the evolving and personal nature of how vibe coding is defined.

\subsubsection{How does the performative aspect of vibe coding on streaming platforms influence the practice?}
\label{sec:results-performance}
Content creation for streaming platforms (YouTube and Twitch) impacts how vibe coding is presented and practiced. The observations below are important to bear in mind when interpreting our results, as while we are attempting to characterise the phenomenon of vibe coding in general, an important aspect potentially affecting the generalisability of our findings is the fact that the videos in our dataset were \emph{performed} for an online audience, which is not a property likely to hold for most vibe coding.

Creators frequently emphasise the impressive capabilities of AI through enthusiastic reactions, with YT15's programmer repeatedly saying ``wow'', such as when the AI-generated code for animations worked correctly. In YT21, the creator explicitly highlights that ``because of the capabilities of AI and Cursor, [the creator] was able to accomplish [building the app],''. Creators emphasise the ease and speed of vibe coding (e.g. ``you can develop this code even without touching a keyboard'', ``boost my productivity''), invoke rhetorical flourishes like ``trust the vibe'' and call the results ``awesome'' or ``incredible''. Even obvious failures may be framed positively (e.g., in one instance a creator hypothesised that transcription errors force the model to ``think more broadly'', becoming a potential asset). These reactions amplify the perception of AI effectiveness for audiences, potentially to the point of exaggeration.

Streaming workflows introduces unique dynamics, as seen in TW1, where the creator frequently responds to chat questions and comments which can cause tangents commonly found in streaming workflows \citep{StreamersTeaching2021} including playing a game on their website because a viewer mentioned wanting an specific achievement. These interactive elements generate spontaneous content that wouldn't occur in private coding but can yield valuable insights, as when they explained that ``the context window is big now so [they don't] have to worry about [token usage] as much'' in response to a viewer's concern about instruction length.

The performance context sometimes encourages procedural shortcuts for narrative flow, as seen in YT21 where the programmer states, ``I could test this API, but let's just wing it,'' demonstrating a desire to maintain video momentum. Similarly, YT22's creator structures development around video-friendly milestones, declaring ``phase one is successfully done'' at a suitable point for a video segment and explicitly mentioning video length as the reason for stopping at certain points, artificially dividing the development process to align with content creation needs.

Despite the emphasis on showcasing AI capabilities, the performances still demonstrate that realistic workflows include debugging and manual intervention (as elaborated in Sections~\ref{sec:debugging} and~\ref{sec:transitions}). Moreover, creators often verbalise their understanding and expectations when they are uncertain. Thus, this running commentary might not just be purely directed at impression management (discussed later in Section~\ref{sec:discussion}) but also a way for the creator to maintain a sense of control and momentum during development. By narrating the process, they might be developing and iterating on their own mental model.

\section{Discussion}
\label{sec:discussion}

\subsection{How vibe coding differs from previous generations of AI-assisted programming}

The initial phase of AI-assisted programming tools, exemplified by early analyses of tools like GitHub Copilot, marked a significant shift in how programmers interact with code generation. Research from this period, such as the studies by \cite{sarkar2022programmingai}, \citet{vaithilingam2022expectation} and \citet{barke2023grounded}, documented how the programmer's experience changes in response to AI capabilities like code completion based on context and comments. The subsequent generation of tools expanded upon these capabilities by evolving the scope and size of code generation within a codebase, and allowing the language model to access other tools that expand the scope of their operations (e.g., running terminal commands to create files, start servers, install packages, etc.). This shift in capabilities is sometimes referred to as making these tools more ``agentic''. Correspondingly, the experience of vibe coding, as observed in our videos, builds upon the experience of prior generations of AI-assisted programming, but introduces distinct nuances in philosophy, workflow, and the programmer's engagement with the AI.

A primary difference lies in the philosophy guiding the interaction. Vibe coding often embraces a more trusting, hands-off approach, prioritising flow or ``vibe'' over strict control. The explicit inspiration from the Karpathy canon encourages relying on the LLM to autonomously handle errors and work around difficult problems, moving away from traditional debugging methods. This introduces a workflow where the programmer may intentionally allow the AI to ``drive'' more consistently, even for familiar tasks, creating an fluid interaction style that is distinct from the segmented modes previously identified \citep{barke2023grounded}.

Prompting takes on a new character in vibe coding. Earlier research noted the difficulties users faced in effectively communicating intent and matching their abstraction level to the model's capabilities, often relying on explicit comments \citep{sarkar2022programmingai,barke2023grounded}. Vibe coding appears to complicate this, featuring prompts that are often significantly longer and exhibit extreme mixing of granularity. Vibe coding prompts can blend vague aesthetic desires, broad ideological goals (like ``make it 10X better''), specific technical constraints (like using a particular library or file location), and may even incorporate direct code snippets or external documentation links for context. Early tools primarily relied on comments or explicit pop-up input fields for prompting \citep{barke2023grounded}, while vibe coding integrates voice, text, and potentially other modalities in a fluid, less structured manner.

The process of evaluation and debugging also shows differences. While earlier AI-assisted programming required validation through examination, execution, documentation lookup, and static analysis \citep{sarkar2022programmingai,barke2023grounded}, vibe coding prioritises rapid, targeted inspection. As detailed in Section~\ref{sec:debugging-reading}, programmers quickly scan code diffs, looking for familiar patterns or keywords (eyeballing, glancing) rather than performing line-by-line reviews. Crucially, traditional manual debugging skills remain essential in vibe coding. When errors occur, vibe coders frequently revert to conventional methods: analysing error messages, using browser developer tools or the terminal, and forming their own hypotheses about the bug and its solution. The model might then be tasked with implementing the specific fix identified by the programmer, although sometimes raw error messages are simply provided to the model for repair. This workflow where the human diagnoses and plans, and AI executes the fix, contrasts with the user experience of understanding and repairing AI code noted in earlier studies \citep{vaithilingam2022expectation}. It also differs from the explicit validation strategies like detailed examination and documentation lookup observed more prominently in the exploration mode of earlier tools \citep{barke2023grounded}.

The programmer's expertise is critical in both paradigms, but its application is different in vibe coding. Early studies found that experienced users spent more time in ``acceleration'' (generating code to fulfil a well-formed intention) \citep{barke2023grounded}. In vibe coding, expertise is vital not just for coding but for strategically interacting with the AI and managing the broader development environment. Experts apply their knowledge to select tools and models, configure projects, perform rapid visual code assessments, manually diagnose complex bugs when the AI fails, and strategically decide when to switch between AI-driven and manual work. They can quickly identify redundant or unwanted AI output and synthesise information across multiple files or external tools like documentation or terminal output. This level of expertise is less about writing code line-by-line (as might be accelerated by earlier tools) and more about orchestrating the model and other tools within a complex development environment (what \citet{lee2025aisurvey} term ``task stewardship'').

Vibe coding is thus an evolution of first-generation AI-assisted programming, that leans into the conversational and generative power of large language models, while still requiring significant human expertise and judgment. It represents a move from code-generating models as an advanced autocomplete or search tool to a more integrated and capable development tool. Previous work had already identified the shortcomings of analogies between AI-assisted programming and programming via search-and-reuse, or compilation, or programming by specification \citep{sarkar2022programmingai}; these analogies are even weaker in the case of vibe coding.

\subsection{A gestalt theory of vibe}

What is the nature and function of the ``vibe'' in vibe coding? This contemporary colloquialism is a word meaning ``the mood of a place, situation, person, etc. and the way that they make you feel'',\footnote{\url{https://dictionary.cambridge.org/dictionary/english/vibe}} ``a distinctive feeling or quality capable of being sensed''.\footnote{\url{https://www.merriam-webster.com/dictionary/vibe}}

We speculatively posit a connection between the vibe of vibe coding to Gestalt psychology (e.g., \citet{gestalt2013}) by focusing on the holistic perception and emergent understanding that characterise both. Vibe coding, in its rapid, iterative nature, encourages a programmer to perceive the AI-generated commentary, code, and agentic actions taken within the IDE as a whole, relying on a continuous ``vibe check'' that corresponds to the gestalt principle that sensory experience of the world is structured as organised wholes (as opposed to parts).

In vibe coding, prompts are often high-level and describe the desired outcome rather than the specific implementation (shifting the focus of programmer intention and articulation towards the whole rather than the parts that compose it). The model then generates code, which, in conjunction with the running application and the context momentum (Section~\ref{sec:context-momentum}) that builds through iteration, can be seen as an emergent gestalt.

The practice of quickly scanning and evaluating AI-generated code can potentially be accounted for through the law of prägnanz (good gestalt), which states that people tend to interpret their sensory experience using heuristics that render their experiences as structured, regular, orderly, symmetrical, and simple. A positive vibe suggests a well-formed and understandable gestalt. Conversely, a negative vibe might signal a lack of coherence or unexpected elements. As we have seen, vibe coding still requires significant human expertise, manual intervention, and critical evaluation of both code and results, especially when the vibes are decidedly off.

As detailed in Section~\ref{sec:debugging-reading}, programmers spend less time deeply understanding every line generated by the model and more time validating its output against their high-level expectations and mental model. The ability to do this appears to be profoundly dependent on the programmer's existing expertise and well-established mental schemas of coding patterns, frameworks, and their specific codebase. This expertise goes beyond the expertise required to write code manually. We posit that this is expertise in understanding the gestalt of \emph{the code within the context of the conversational session} with the model. Programmers repeatedly exhibited the ability to look at a block of code and immediately identify its purpose within the overall program, and identify potential problems or inefficiencies quickly. This is possible because the code conforms to a mental schema that the programmer possesses, and is capable of rapidly updating as the code evolves, so that the vibe gestalt continues to be useful for future steering and verification. This expert schema, unlike in traditional programming, goes beyond the well-documented expert ability to read and synthesise a code base quickly, because it incorporates several novel constituent elements in the gestalt: the AI-generated commentary, the conversational chat history and context momentum, and knowledge and prior expectations about model capabilities.

\subsection{The consequences of material disengagement from code}

Vibe coding workflows exhibit the phenomenon of material disengagement from the traditional material substrate of programming: code. Programmers step back from directly manipulating the code itself, instead using AI tools as intermediaries to generate and modify large sections of the codebase. AI assumes the role of handling the tedious material manipulation of code. This fundamentally changes the programmer's relationship with the textual material of the program.

Vibe coding represents a reorientation of material engagement towards the AI tool itself as a mediating entity. Importantly, the AI's generated code, commentary, and errors actively facilitate the formation and refinement of the programmer's intentions, which evolve iteratively from initial ideas as the session progresses based on this dialogue.

This can be understood in terms of Material Engagement Theory (MET) \citep{malafouris2019mind}, which posits that the mind is not an isolated internal entity but is fundamentally constituted through its engagement with the material world. Thinking, in this view, is often best understood as ``thinging'': a process of thinking primarily with and through things, where the reciprocal interaction between mind and material actively shapes cognition and the development of skill and intention. Examples like pottery making illustrate this, where the interaction with the resistances and affordances of clay is inseparable from the potter's thinking and the development of skill and intention.

From the MET perspective, the AI tool within the vibe coding environment could be interpreted as a new form of ``thing'' or mediating material. The programmer's cognitive process would thus be enacted not by manipulating code syntax, but by engaging with the AI's interface, formulating prompts, and evaluating the AI's material output (the generated code and its behaviour). This interpretation of AI-as-thing suggests that vibe coding involves a shift in the object of material engagement (from code to AI) rather than its complete abandonment, with the programmer's expertise now oriented towards navigating this new material substrate.

What is potentially lost in this disengagement from direct code manipulation, of course, is the deep, enactive understanding and skill formation that arises from grappling firsthand with the material resistances inherent in code's syntax, structure, and debugging challenges in a manual workflow. Dialogue with the code's own constraints and affordances (indeed, the cognitive \emph{dimensions} that arise as a consequence of manipulating a \emph{notation}), like the potter's interaction with clay, is diminished or altered when mediated by an AI agent.

Nevertheless, the iteration central to vibe coding demonstrates how engagement with the model's material responses acts as a form of resistance that actively shapes and refines the programmer's intentions and strategies. The unexpected outputs, errors, or failures to meet requirements function as the resistances or affordances of this new mediating material, pushing back against the programmer's initial intent and necessitating adjustments or refinements. While vibe coding workflows often seek to minimise friction, the inherent pushback from the model's outputs, and the programmer's active evaluation and response to these, constitutes a form of material engagement that shapes intentions.

This dynamic aligns with the concept of designing ``productive resistances'' \citep{sarkar2024intention} in AI tools to cultivate intention, counteracting tendencies towards weakened intentionality, and thereby mechanised convergence \citep{sarkar2023aiknowledgework} of outcomes. Intentionally designed productive resistances could formalise and enhance a process already occurring organically within vibe coding workflows.

Moreover, our analysis finds that code itself does not cease to be a material substrate in vibe coding. Programmers still engage with the code material, but the nature of this engagement shifts profoundly, involving new material manoeuvres: from rapid, high-level review, scanning diffs, identifying patterns, relying on gestalt properties and visual cues, to inspecting keywords and structure to verify correctness or diagnose issues, and indeed to direct, manual writing, editing, and debugging (albeit tactically and selectively, rather than as the norm). Crucially, as we have abundantly seen, navigating the vibe coding workflow still demands substantial expertise in manipulating the underlying code material to engage in these ways.

Interestingly, an example from YT21 sheds light on yet another complexity of how AI might change a programmer's relationship with the materiality of code. Specifically, because the programmer is using AI, they seem to be able to use more ``vanilla'' JavaScript and rely less on higher-level libraries like React, or abstraction-rich languages such as TypeScript. High-level libraries and languages, are, of course, abstractions created in direct response to the perceived limitations of working in a particular material substrate. Research on earlier generations of AI-assisted programming has noted how the human expertise embodied in such abstractions was a key contributor to the successful application of earlier models, which generated a relatively small number of lines of code at a time \citep{sarkar2022programmingai}, compared to the sweeping, cross-codebase changes made by LLM agents during vibe coding.

It is worth dwelling for a moment here on the implications of this example for manual programming versus AI-assisted programming. For a manual programmer, high-level libraries are valuable because they reduce the material barriers for engaging with code and are thus important for their workflow. For a manual programmer, writing in vanilla JavaScript can be verbose, tedious, and error-prone. And for an AI-assisted programmer working with a prior generation of AI code generation tools, the expressiveness of high-level libraries greatly amplified what could be achieved within the small contexts and size of output that those models were capable of.

However, when working with contemporary agentic AI tools, the code-generating model can handle this tedious material manipulation of the low-level abstractions while maintaining enough context to build more complex, ad-hoc abstractions as necessary. This might enable programmers to regain the benefits of working with lower-level code -- principally, to avoid external dependencies and thus retain greater control and flexibility over their own codebase. Paradoxically, this suggests that the flight from material engagement in vibe coding may also enable a return to working effectively with lower material substrates.

The challenge (and at this point, this is hardly a novel observation) lies in balancing the benefits of offloading tedious manipulation against the potential loss of the distinct form of cognitive engagement and skill development that arises from directly confronting and resolving the intrinsic resistances posed by the material properties of code itself.

\paragraph{Vibing beyond coding.}
As touched upon in Section~\ref{sec:introduction}, the vibe coding workflow, where the user avoids direct manipulation of the ``raw'' material substrate (e.g., text in a document, formulas in a spreadsheet, paint on a canvas) may spread to other AI-assisted knowledge workflows. However, the potential for turning these workflows into ``vibe'' versions of themselves may differ between domains. The programming context has resources that the practitioner can draw upon to support their understanding of the vibe gestalt that do not easily transfer to other domains, such as the saliency of identifiers and component structure. Code can be tested through unit tests or throw formal exceptions to aid detection of errors. Code can also be compiled and deployed for visual and interactive inspection by the user to judge whether code generation has been satisfactory. 

Text, on the other hand, is altogether more complex. Auditing generated text may inevitably require line-by-line reading to ensure that a ``vibe written'' document is accurate and reflects the user's intent. Therefore, expanding vibe coding workflows to other domains may require creating structural support for users to co-audit \citep{gordon2024coaudit} AI-intermediated workflows  before we see vibing effectively expand to domains beyond coding.

\subsection{Impression management and self-presentation: vibing for an audience}
\label{sec:discussion-performance}

The practice of impression management involves the strategic curation of information within social encounters, enabling individuals to consciously or unconsciously direct how observers form judgments about people, objects, or situations. It is sometimes seen as synonymous with self-presentation, which is aiming to influence the perception of one's image \citep{Goffman1956} (though it is straightforward to see that self-presentation is a subset of impression management). We find that creators engage in impression management: over and above the typical concerns of maintaining viewer engagement and satisfaction on streaming platforms, they act specifically to manage perceptions of their expertise and value in the context of contemporary social discourse around AI use in knowledge work.

To understand how our participants engage in impression management, it is necessary to contextualise our observations within the emerging understanding of the social dynamics of AI use. \citet{reif2025evidence} provides empirical evidence for a social evaluation penalty associated with using AI tools in the workplace. Their studies show that people anticipate and receive negative judgments from others, being perceived as lazier, less competent, and less diligent than those who use non-AI tools or no help at all. This penalty is linked to observers making negative dispositional inferences, attributing the use of AI to personal deficits rather than situational factors. There is thus a dilemma where the productivity benefits of AI can come at a social cost, leading some employees to conceal their AI use.

Adding to this, \citet{SCHILKE2025104405} describe the ``transparency dilemma'' of AI use. Their experimental evidence demonstrates that disclosing the usage of AI compromises trust in the user. This occurs because AI disclosure reduces perceptions of legitimacy. In many work contexts, there is a normative expectation that outputs should be the result of human expertise and judgment, and disclosing AI involvement is perceived as a deviation from these norms, undermining the legitimacy of the work process. Paradoxically, people who try to be trustworthy by transparently disclosing AI usage are trusted less. While disclosure erodes trust, being exposed for using AI by a third party has an even more detrimental effect on trust.

\citet{sarkar2025aishaming} offers a theoretical account for these negative social evaluations, framing AI shaming as a form of boundary work driven by class anxiety among knowledge workers. This perspective suggests that negative judgments and shaming of AI use arise from a perceived threat to the identity, value, and exclusivity of knowledge work professions. AI is seen as potentially eroding the ``moat'' of ``extensive preparation'' that historically protected these roles. Arguments used in AI shaming often question the quality, creativity, and legitimacy of AI-generated work, sometimes portraying AI users as ``tasteless novices'' or lacking the necessary skill, effort, or struggle associated with traditional knowledge work.

Vibe coding creators deploy impression management to directly confront this negative social landscape. Instead of concealing their AI use to avoid negative judgments, the video creators make their AI use central to their performance. They understand the potential for viewers to make negative dispositional inferences or engage in shaming by perceiving their AI use as evidence of laziness, lack of skill, or merely producing ``slop''. To counteract this, they strategically demonstrate their technical expertise, debugging skills, and ability to guide the model. By showing they are in control and possess the knowledge to validate and correct the AI's output, they differentiate themselves from the ``unskilled opportunists'' or ``uncultivated dilettantes'' \citep{sarkar2025aishaming} that the shaming discourse might target. They thus project an image of competence and expertise, even when relying on AI. This enables them to maintain their professional image and credibility while openly embracing AI. They are, in essence, performing a positive form of boundary work, attempting to define skilled AI-assisted programming as a legitimate and even advanced form of knowledge work.

\subsection{Limitations}

We note the novelty of vibe coding, which can be said to have begun in earnest in February 2025 with the publication of the Karpathy canon. This meant that our corpus of videos was limited to an initial \textasciitilde35 hours of vibe coding content, \textasciitilde8.5 hours of which represented high quality videos that we analysed. This is unquestionably quite a small corpus, and future work is necessary to expand our analysis. Moreover, as outlined from the outset in Section~\ref{sec:introduction}, vibe coding is a rapidly evolving, emerging, and negotiated practice, which means that our data bears witness to vibe coders merely \emph{beginning} to undertake such exploration and negotiation. There is much still to learn about how the practice changes as vibe coders gain experience and develop new rhythms in the vibe coding workflow and as IDEs evolve to support them.

While we have discussed the performative aspect of these videos in detail (Sections~\ref{sec:results-performance} and~\ref{sec:discussion-performance}), it must be acknowledged that performance aspects may skew workflows in ways that do not accurately reflect vibe coding in other realistic scenarios. To remedy this, in future work, it will be necessary to study vibe coding through other data sources, such as through laboratory experiments, or interviews and diary studies.

Finally, none of the selected videos depicted non-expert end-users (in particular, non-programmers). So, while our findings show how experienced coders can apply their expertise to address many of the challenges of vibe coding workflows, we do not know how non-programmers might meet these challenges and how they create barriers to effective vibe coding. Critically, we have studied the role of expertise in vibe coding in some detail (e.g., Section~\ref{sec:results-expertise}), and concluded that programming expertise is not only important, but essential for successful vibe coding. But we can only claim that this is true insofar as the vibe coder is in possession of said expertise, because we have not studied non-experts. Thus, research is needed to understand what kind of vibe coding -- perhaps an altogether different activity, despite bearing superficial resemblances -- might be practiced by non-experts.

\section{Conclusion}
\label{sec:conclusion}

This study is the first empirical analysis of vibe coding, an emerging programming paradigm where developers primarily author and edit code by interacting with code-generating AI through natural language prompts rather than direct code manipulation. Through framework analysis of curated think-aloud videos from YouTube and Twitch, we examined how programmers form goals, conduct workflows, and deploy expertise when engaging in this novel form of programming.

We find that vibe coding represents a meaningful evolution of traditional AI-assisted programming, characterised by episodes of iterative goal satisfaction where developers cycle through prompting, evaluation, debugging, and refinement. The workflow consistently involves strategic delegation of code generation to AI while maintaining human oversight through rapid evaluation and targeted intervention. Effective vibe coders apply varied and blended prompting strategies, ranging from high-level, diffuse, and subjective directives to detailed, fine-grained, and technical specifications.

Our observations challenge the notion that vibe coding eliminates the need for programming expertise. Instead, we observe a redistribution of expertise deployment. Traditional coding knowledge is redirected toward prompt and context management, rapid code evaluation, bug identification and resolution, and decisions about when to transition between AI assistance and manual intervention. Trust in AI tools is therefore granular, dynamic, and contingent, developed through iterative verification, and not blanket acceptance.

These findings may have broader implications for knowledge work. Vibe coding represents an early manifestation of material disengagement, where practitioners orchestrate content production through AI intermediaries rather than direct manipulation. However, our analysis suggests that such workflows still require substantial expertise in the underlying material substrate (in this case, code) to navigate effectively.

As vibe coding practices continue evolving, this research provides a foundational understanding for future investigations into human-AI workflows in programming and knowledge work more broadly.

\bibliography{references}
\bibliographystyle{apalike}

\begin{appendices}
\section{Video sources}
\label{apx:video-list}

This table lists the videos included in our analysis. Our use of videos in the various research phases are indicated as follows under the ``Research phases'' column: ``sensitising'' means that 10-20 minute excerpts of the video were viewed during the initial sensitising phase; ``development'' means that the entire video was viewed in detail by both researchers with a view to developing the categories for the framework and the criteria for selecting the final videos to analyse; ``analysis'' means the final framework analysis was applied to this entire video.

\begin{footnotesize}
\begin{longtable}{@{}lp{5cm}lll@{}}
\caption{Sources Table}\\
\toprule
ID & URL and video title & Date published & Duration & Research phases \\
\midrule
\endfirsthead

\caption[]{Sources Table (continued)}\\
\toprule
ID & URL and video title & Date published & Duration & Research phases \\
\midrule
\endhead

\midrule
\multicolumn{5}{r@{}}{{Continued on next page}} \\
\endfoot

\bottomrule
\endlastfoot

YT1 & \url{https://www.youtube.com/watch?v=5k2-NOh2tk0} \newline What is ``Vibe Coding''? Here's how I do it... & 23-Feb-25 & 00:17:19 & Sensitising \\
YT2 & \url{https://www.youtube.com/watch?v=z7aZBxjfDIw} \newline Build iOS apps with 0 Knowledge | Vibe Coding | TinderTodo App | Swift & 16-Mar-25 & 00:26:22 & Sensitising \\
TW1 & \url{https://www.twitch.tv/acorn1010/video/2398186371} \newline !foony | VIP | Vibe Coding | !idea & 06-Mar-25 & 05:30:50 & Sensitising, analysis \\
YT3 & \url{https://www.youtube.com/watch?v=OVnDfcRtFew} \newline Vibe Coding: Email Campaigns \& Agents | Chatbot Builder AI & 14-Mar-25 & 01:53:49 & Sensitising \\
YT4 & \url{https://www.youtube.com/watch?v=YWwS911iLhg} \newline Vibe Coding Tutorial and Best Practices (Cursor / Windsurf) & 05-Mar-25 & 00:21:48 & Sensitising \\
YT5 & \url{https://www.youtube.com/watch?v=k9WrU6uR2YM} \newline Vibe Coding my Entire Landing Page with AI (Lovable): & 11-Mar-25 & 00:11:11 & Sensitising, development \\
YT6 & \url{https://www.youtube.com/watch?v=xh5PhzZZcnQ} \newline Vibe Coding with Wispr Flow and Cursor & 12-Mar-25 & 00:09:49 & Sensitising \\
YT7 & \url{https://www.youtube.com/watch?v=dan3QfN3CDU} \newline Karpathy Vibe Coding Full Tutorial with Cursor (Zero Coding) & 06-Feb-25 & 00:33:47 & Sensitising \\
YT8 & \url{https://www.youtube.com/watch?v=g84CGmelvSU} \newline Vibe Coding: Launch Your SaaS with AI (Cursor, Supabase, \& Stripe) & 28-Feb-25 & 00:02:35 & Sensitising \\
YT9 & \url{https://www.youtube.com/watch?v=GwhFjhMF6OA} \newline Vibe coding using ChatGPT & 06-Feb-25 & 00:10:56 & Sensitising \\
YT10 & \url{https://www.youtube.com/watch?v=ulJrdYXLo9I} \newline Vibe Coding a FULL Game ?? (AI Coding) & 04-Mar-25 & 00:11:52 & Sensitising \\
YT11 & \url{https://www.youtube.com/watch?v=faezjTHA5SU} \newline Complete Guide to Cursor For Non-Coders (Vibe Coding 101) & 11-Feb-25 & 02:28:16 & Sensitising \\
YT12 & \url{https://www.youtube.com/watch?v=JEU07S2WyDs} \newline Vibe Coding in 2025: No Typing, No Stress, Just AI! & 08-Mar-25 & 00:21:03 & Sensitising, development \\
YT13 & \url{https://www.youtube.com/watch?v=i0gWDz9EUgI} \newline Vibe Coding 101 & 11-Feb-25 & 00:13:48 & Sensitising \\
YT14 & \url{https://www.youtube.com/watch?v=yGZtu_6E1l8} \newline Vibe Coding in Databutton AI - FREE AI Code Editor for Full stack development in Claude Sonnet 3.7 & 03-Mar-25 & 00:10:22 & Sensitising \\
YT15 & \url{https://www.youtube.com/watch?v=irBnUAq3MAw} \newline VIBE CODING with CURSOR - Draw w/ HTML and JS & 17-Feb-25 & 00:17:45 & Sensitising, development, analysis \\
TW2 & \url{https://www.twitch.tv/coolaj86/video/2399572024} \newline Vibe Coding Hangout: Live w/ Grok3 Unfurl-as-a-Service:Link Unshortener & 07-Mar-25 & 05:57:50 & Sensitising \\
TW3 & \url{https://www.twitch.tv/vadimnotjustdev/video/2402985837} \newline Vibe Coding a React Native Game & 11-Mar-25 & 02:41:35 & Sensitising \\
YT16 & \url{https://www.youtube.com/watch?v=HBwXs99LFlw} \newline VIBE CODING 3 min demo | Cursor + o3-mini + SuperWhisper & 15-Feb-25 & 00:03:20 & Sensitising \\
YT17 & \url{https://www.youtube.com/watch?v=Nt2Lkdy3f5Y} \newline ?Vibe? Coding AI Agent to do sales for me (OpenAI Agents SDK) & 17-Mar-25 & 00:40:04 & Sensitising \\
YT18 & \url{https://www.youtube.com/watch?v=_QOvocOFLbo} \newline Vibe Coding - Code \& Chill & 16-Mar-25 & 00:22:28 & Sensitising, development \\
YT19 & \url{https://www.youtube.com/watch?v=TwSYePsdfOk} \newline WIP at Recall: Vibe Coding Crypto Alpha Detection Agents \#1 & 15-Mar-25 & 00:27:28 & Sensitising \\
YT20 & \url{https://www.youtube.com/watch?v=Pe8ghwTMFlg} \newline ?? VS Code - Agent Mode UPGRADE! & 07-Mar-25 & 01:32:45 & Sensitising \\
YT21 & \url{https://www.youtube.com/watch?v=EAPWrpvIOjs} \newline Vibe Coding with AI: Watch me fix a chat interface in real-time in this long video & 06-Feb-25 & 01:29:37 & Sensitising, analysis \\
YT22 & \url{https://www.youtube.com/watch?v=SNARzs6jzWY} \newline Vibe Coding Startup Using Most Advanced AI — Phase 1 & 14-Mar-25 & 00:32:05 & Sensitising, analysis \\
YT22b & \url{https://www.youtube.com/watch?v=u4yqSsYGpcc} \newline Vibe Coding Startup: Dashboard — Phase 2 Part 1 & 21-Mar-25 & 00:37:02 & Sensitising, analysis \\
YT23 & \url{https://www.youtube.com/watch?v=DvDOb0ZezQQ} \newline Vibe coding actually sucks & 19-Mar-25 & 00:28:31 & Sensitising \\
YT24 & \url{https://www.youtube.com/watch?v=NYVaCr3T1T0} \newline Vibe Coding is Actually INSANE... (Vibe Coding Tutorial for Beginners) & 21-Mar-25 & 00:38:24 & Sensitising \\
YT25 & \url{https://www.youtube.com/watch?v=opi1s_5Dm-c} \newline He makes \$750 a day `Vibe Coding' Apps (using Replit, ChatGPT, Upwork) & 21-Mar-25 & 00:45:16 & Sensitising \\
YT26 & \url{https://www.youtube.com/watch?v=faPSZV5XwyI} \newline Start Vibe Coding Like a Pro, Here's How & 17-Mar-25 & 00:21:48 & Sensitising \\
YT27 & \url{https://www.youtube.com/watch?v=y3vQBBmE3Cc} \newline AI Created My Game in 6 Hours... Vibe Coding so you don't have to & 20-Mar-25 & 00:08:31 & Sensitising \\
YT28 & \url{https://www.youtube.com/watch?v=icRXv9RXhXI} \newline Vibe Coding FULL Course + WIN MacBook Pro, PlayStation 5 ?? & 27-Mar-25 & 01:22:39 & Sensitising \\
YT29 & \url{https://www.youtube.com/watch?v=jTaqixu79qU} \newline Vibe Code SaaS \& Mobile Games (Grok, Bolt, Cursor, Prompts) & 18-Mar-25 & 00:16:28 & Sensitising \\
TW4 & \url{https://www.twitch.tv/codesinthedark/video/2414451732} \newline DAY ??x3 - Vibe coding a game using Cursor \#vibejam & 24-Mar-25 & 01:32:40 & Sensitising \\
YT30 & \url{https://www.youtube.com/watch?v=OZaxtm3RyCw} \newline Vibe Coding Made Easy: The Essential Cursor AI Tool! & 30-Mar-25 & 00:12:16 & Sensitising \\
YT31 & \url{https://www.youtube.com/watch?v=y9XEBnNvu2Q} \newline Vibe Coding and Portfolio Reviews & 28-Mar-25 & 01:12:38 & Sensitising \\
YT32 & \url{https://www.youtube.com/watch?v=_yKDiRlToSs} \newline Vibe Coding For Non Coders - I built an online game in 30 seconds using AI & 25-Mar-25 & 00:05:19 & Sensitising \\
YT33 & \url{https://www.youtube.com/watch?v=5qwucCaHpWY} \newline BUILDING A GAME IN 7 DAYS & 21-Mar-25 & 00:12:16 & Sensitising \\
YT34 & \url{https://www.youtube.com/watch?v=78jina4V7j4} \newline One Shotting a Rank and Rent Site With Lovable and Going Live (Vibe Coding With Alex) & 02-Apr-25 & 00:26:16 & Sensitising \\
YT35 & \url{https://www.youtube.com/watch?v=X-xiJgkqnok} \newline Episode \#503 - Vibe Coding & 01-Apr-25 & 00:17:33 & Sensitising \\
\end{longtable}
\end{footnotesize}

\section{Analysis framework}
\label{apx:framework}

\begin{footnotesize}
\begin{longtable}[tbp!]{|p{3cm}|p{3cm}|p{9cm}|}
\caption{Qualitative analysis framework for vibe coding think aloud videos}
\label{tab:framework}
\\\hline
\textbf{Top Level Category} & \textbf{Subcategories} & \textbf{Elaboration} \\
\hline
\endfirsthead
\hline
\textbf{Top Level Category} & \textbf{Subcategories} & \textbf{Elaboration} \\
\hline
\endhead
\hline
\endfoot
\hline
\endlastfoot
Goals & a. What do they build? & i. What applications did vibe coders build? \\
 &  & ii. What types of applications are created (e.g., landing pages, games, etc.)? \\
 &  & iii. Is vibe coding fully appropriate to complete a project, or does it require transitions to manual coding for more complex projects like games? \\
\cline{2-3}
 & b. Expectation – for full/partial success, or just exploration? & i. Do people go into vibe coding expecting to fully succeed, or are they taking an exploratory attitude? \\
 &  & ii. Are users generally expecting complete success with vibe coding, or do they experiment to see how far the technology can be pushed? \\
\hline
Intentions & a. Initial intention & i. How is the intention for an app formed and refined in vibe coding? \\
 &  & ii. Is the intent pre-formed before coding begins, or does it evolve through iterative interactions with the AI? \\
 &  & iii. Which parts of the intent are established beforehand, and how does intent spread across various prompts and phases? \\
\cline{2-3}
 & b. Is intention refined? How and why? & i. What AI workflow resources (e.g., exploration, seeking alternatives) support and refine this intent? \\
 &  & ii. Does the AI itself contribute to shaping the final intent by filling in gaps? How does the back-and-forth interaction with the AI shape the final output? \\
 &  & iii. How does the iterative and conversational nature of vibe coding affect users’ creativity and problem-solving strategies? \\
 &  & iv. Does this process encourage more exploration or lead to a reliance on the AI’s suggestions? \\
\hline
Workflow & a. Stages of workflow & i. What are the stages of the vibe coding workflow? \\
 &  & ii. How is time and effort allocated across different stages (e.g., programming, fault localization, debugging versus planning and testing)? \\
 &  & iii. How do these stages compare with those in more traditional AI-enhanced programming workflows and prior identified AI workflows such as iterative goal satisfaction? \\
\cline{2-3}
 & b. Portfolio of tools and technologies & i. How do different tools work together in the context of vibe coding? \\
 &  & ii. How are other resources (such as documentation, web searches, etc.) integrated into the vibe coding workflow? \\
\hline
Prompting & a. Any setup work? & i. How do creators set up a vibe coding session (e.g., editing prompt instructions vs. using default settings)? \\
\cline{2-3}
 & b. Explicit prompting strategies & i. What are the prompting strategies in vibe coding? \\
 &  & ii. What methods are employed to initiate and guide the AI during the coding process? \\
\cline{2-3}
 & c. Granularity of prompts – high/low level & i. What is the granularity of prompts and iterations? \\
 &  & ii. How fine-grained are the iterative steps in the vibe coding process? \\
 &  & iii. How is aesthetic intent communicated through natural language prompts translated (or misinterpreted) by generative AI into technical specifications and code? \\
\cline{2-3}
 & d. Single/multi-objective prompts & i. Do users address multiple issues at once or handle one problem at a time? \\
\cline{2-3}
 & e. What prompting modes are used – voice/text/images? & i. What is the role of different modes (voice, text, screenshot, etc.) in vibe coding? \\
 &  & ii. How do creators incorporate various input modes into the vibe coding process? \\
 &  & iii. How is code reuse managed, for instance by borrowing online examples to steer the AI? \\
\hline
Debugging & a. What debugging strategies are applied? & i. What debugging strategies are employed in vibe coding? \\
 &  & ii. How do creators approach debugging code generated through vibe coding? \\
\hline
Challenges & a. What challenges (besides code bugs) are encountered in achieving goals? & i. What challenges and barriers are encountered during vibe coding? \\
 &  & ii. What technical, conceptual, or workflow-related obstacles arise throughout the process? \\
\cline{2-3}
 & b. How are they addressed? & i. What strategies are adopted to address these challenges? \\
\hline
Expertise & a. When is expertise used? & i. How do experts deploy their expertise in vibe coding? \\
 &  & ii. In what ways do experts deploy their expertise? \\
\cline{2-3}
 & b. What kinds of expertise are used? & i. Is the relevant expertise strictly developer knowledge, or does it include understanding system requirements, end-to-end design, and the broader technical ecosystem? \\
 &  & ii. Could a tech-forward attitude be as significant as traditional coding skills in this context? \\
\cline{2-3}
 & c. When do they switch to manual work? & i. When do practitioners transition from vibe coding to manual editing (code, text, image, artifact)? \\
 &  & ii. At what point in the workflow is it appropriate to jump into traditional manual editing? \\
\hline
Trust & a. Any reflections on trust, overtrust, reliance? & i. What is the nature of trust in vibe coding? \\
 &  & ii. How do users develop trust in AI-generated outputs, and what are the risks of overtrust or overreliance on the technology? \\
\hline
Definition of vibe coding and performance & a. How do programmers define vibe coding? & i. When is an activity considered vibe coding versus simply using AI tools? \\
 &  & ii. Does vibe coding include not only code generation but also debugging, planning, and testing? \\
 &  & iii. Where does vibe coding fit on the spectrum of AI-assisted programming activities? \\
\cline{2-3}
 & b. How does performance for YouTube/Twitch skew production? & i. How does the performative aspect of vibe coding (e.g., YouTube demonstrations) influence the practice and perception of generative AI in programming? \\
 &  & ii. How might public demonstrations affect tool choice, prompting strategies, and the framing of users’ experiences? \\
\hline
\end{longtable}
\end{footnotesize}

\end{appendices}

\end{document}